\documentclass[12pt]{iopart}
\usepackage{fullpage,graphicx,epsf}
\usepackage{psfrag}
\usepackage{epstopdf}
\begin{document}
\title{Condensation and Extreme Value Statistics} 
\author{Martin R. Evans$^{1,2}$ and Satya N. Majumdar$^{2}$}

\address{
$^1$ SUPA, School of Physics, University of Edinburgh, \\
Mayfield Road, Edinburgh EH9 3JZ, UK.\\
$^2$ Laboratoire de Physique Th\'eorique et Mod\`eles
Statistiques,\\ Universite Paris-Sud, Bat 100, 91405, Orsay-Cedex,
France.}
\ead{m.evans@ed.ac.uk, majumdar@lptms.u-psud.fr}

\begin{abstract}
We study the factorised steady state of a general class of mass transport
models in which mass, a conserved quantity, is transferred stochastically
between sites. Condensation in such models is exhibited when above a critical
mass density the marginal distribution for the mass at a single site develops
a bump, $p_{\rm cond}(m)$, at large mass $m$. This bump corresponds to a
condensate site carrying a finite fraction of the mass in the system. Here, we
study the condensation transition from a different aspect, that of extreme
value statistics. We consider the cumulative distribution of the largest mass
in the system and compute its asymptotic behaviour. We show 3 distinct
behaviours: at subcritical densities the distribution is Gumbel; at the
critical density the distribution is Fr\'echet, and above the critical density a
different distribution emerges. We relate $p_{\rm cond}(m)$ to the probability
density of the largest mass in the system.

\end{abstract}
\date{\today}
\maketitle

\section{Introduction}
Mass transport models form a general class of nonequilibrium systems
where some conserved quantity, which we will refer to as mass,
is transported stochastically
between the sites
of a lattice, according to certain prescribed  dynamical  rules \cite{EMZ04}.
At various levels of description such models may  represent
the dynamics of  traffic flow
\cite{OEC,KMH05}, granular clustering \cite{granular}, 
phase ordering \cite{KLMST},
network rewiring \cite{DM,AELM,AHE06}, force propagation \cite{CLMNW},
aggregation and fragmentation \cite{MKB,ARAP} and  energy transport
\cite{Bertin}
---for a review see
\cite{EH05}.

Of particular interest is the nonequilibrium steady state 
which is attained in the long 
time limit. In general the structure of nonequilibrium steady states is
not known, however in some cases
the steady state factorises. That is,
the joint distribution of the collection of masses $\{ m_i \}$ at sites $i =1, \ldots ,L$ 
is given by
\begin{equation}
P(m_1,\cdots ,m_L)=\frac{\prod_{i=1}^Lf(m_i)}{Z(M,L)}\,\delta \left(
\sum_{j=1}^Lm_j-M\right)   \label{prodm1}
\end{equation}
where
$Z(M,L)$ is just the normalization
\begin{equation}
Z(M,L)=\prod_{i=1}^L\int_0^\infty dm_i\,f(m_i)\delta \left(
\sum_{j=1}^Lm_j-M\right)  \label{Zcan}
\end{equation}
and is the analogue of the `canonical partition function'.
Note that (\ref{prodm1}) is 
a product of single-site weights $f(m_i)$ but
the $\delta$-function in (\ref{prodm1}) imposes the global constraint
that the total mass, $M$, in the system is conserved:
\begin{equation}
M = \sum_{j=1}^Lm_j\;.
\end{equation}
Thus, correlations are induced between sites and in general the 
single-site mass {\em probability distribution}, i.e., the marginal
$p(m,M,L)=\int P(m, m_2,m_3,\ldots,m_L)\,\prod_{i=2}^L dm_i \neq f(m)$.
(For brevity we will often use $p(m)$ as a shorthand for
$p(m,M,L)$.)

In the case of a simple class of one-dimensional asymmetric mass transport
models, a necessary and sufficient condition for factorisation has been
determined. In these models the dynamics is defined as follows. At each time
step, a portion, $\tilde{m}_i\le m_i$ of the mass at each site, is chosen from
a distribution $\phi(\tilde{m_i}|m_i)$ and is transferred to site $i+1$. The
stationary state is factorised provided that the kernel $\phi(\tilde{m}|m)$ is
of the form
\begin{equation}
\phi(\tilde{m}|m) = \frac{u(\tilde{m})v(m-\tilde{m})}{f(m)}\;,
\end{equation}
where $u(z)$ and $v(z)$ are arbitrary non-negative
functions. 
The single-site weight is then given by
\begin{equation}
f(m) = \int_0^m d\tilde{m}\  u(\tilde{m}) v(m-\tilde{m})\;.
\label{ssw}
\end{equation}
This model is general enough to include many well-known
models as special cases \cite{EMZ04}. Choosing the chipping kernel 
$\phi(\tilde{m}|m)$ appropriately, recovers the Zero-Range Process
(discrete, integer valued masses with $\tilde{m}$ restricted to 0,1) which
always has a factorised steady state, and the
Asymmetric Random Average Process (continuous masses 
with $\tilde{m}$ chosen as a random fraction of $m$) 
which exhibits a factorised steady state in certain cases \cite{ARAP,ZS,ZEM04}.
Moreover, the model encompasses both discrete and
continuous time dynamics.  The condition for
factorisation has been generalised to arbitrary dimensions and arbitrary
graphs \cite{EMZ06a}, where generally one is able to demonstrate 
sufficient conditions for factorisation.
We also note that factorised steady states have been extensively studied in
the queueing theory literature, see e.g. \cite{queue}.

By choosing the dynamics appropriately various forms for the steady state
weight $f(m)$ can be produced from (\ref{ssw}). Of particular interest has been
the case where for large $m$
\begin{equation}
f(m) \simeq A {m^{-\gamma}}\;.
\label{fasymp}
\end{equation}
In the following we choose the normalisation constant $A$ so that
$\int_0^{\infty} {\rm d}m\ f(m) =1$. Then, if the index $\gamma>2$ the
phenomenon of condensation occurs
\cite{BBJ,OEC,Evans00,JMP00,GSS03,Godreche03,MEZ05,EMZ06,FLV07}. This is manifested by
three distinct large $m$ behaviours of $p(m)$ as the global mass density,
$\rho$, given by
\begin{equation}
\rho = \frac{M}{L},
\end{equation}
is increased  (see Figure~\ref{pmfig}):
\begin{figure}[htb]
\begin{center}
\includegraphics[width=10cm]{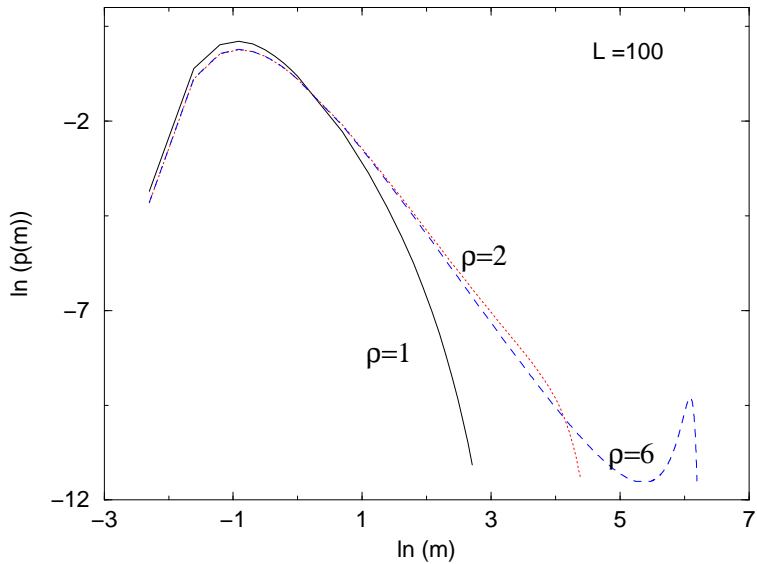}
\caption{\label{pmfig} The exact single-site mass distribution $p(m)$ plotted
  for a particular choice of $f(m)$ with $\gamma = 5/2$ and $\rho_c =2$, and
  system size $L=100$: full line $\rho=1$ (subcritical); dotted line $\rho=2$
  (critical); dashed line $\rho= 6$ (supercritical). The condensate bump,
  $p_{\rm cond}(m)$, is apparent in the supercitical curve.}
\end{center}
\end{figure}

\begin{enumerate}
\item When $\rho< \rho_c$, $p(m) \simeq B f(m) z^m$
where $z<1$ and $B$ is a normalisation constant for $p(m)$. 
This is the fluid phase where $p(m)$ decays exponentially for
large enough $m$
\item When $\rho= \rho_c$, $p(m) \simeq  f(m)\simeq A {m^{-\gamma}}$.
This is the critical point
where $p(m)$ decays  as a power law.
The critical density $\rho_c$ is given by
\begin{equation}
\rho_c = \int_0^\infty {\rm d}m\, m f(m)\;.
\label{rhoc}
\end{equation}
\item When $\rho> \rho_c$, for finite but large $L$, $p(m)$ develops an
  additional bump centred at $m=L(\rho-\rho_c)$. Roughly speaking, for large
  $L$, $p(m) \simeq f(m) + p_{\rm cond}(m,M,L)$. This is the condensed phase
  where this extra piece of $p(m)$, $p_{\rm cond}(m,M,L)$, emerges in addition
  to the critical point distribution. This piece (the bump in $p(m)$) carries
  a weight $1/L$. It thus represents a single site containing the excess mass
  $L(\rho-\rho_c)$, which coexists with a background of power-law distributed
  critical mass. The large system-size behaviour of the bump $p_{\rm
    cond}(m,M,L)$ was determined \cite{MEZ05,EMZ06} and shown to have distinct
  forms in the regimes $\gamma >3$, where on the scale $m - L(\rho-\rho_c) =
  O(L^{1/2})$ it is gaussian, and $2<\gamma<3$, where the distribution is
  highly asymmetric and non-gaussian. These two forms imply that the number of
  particles in the condensate has fluctuations of $O(L^{1/2})$ for $\gamma >3$
  but anomalously large fluctuations of $O(L^{1/(\gamma-1)})$ for $3> \gamma
  >2$.

\end{enumerate}
This description of condensation has been from the point of view of the
marginal distribution $p(m)$ \cite{MEZ05,EMZ06} in the thermodynamic 
(large $L$)  limit.
However,  condensation is a co-operative behaviour resulting in the 
emergence of a single extensive mass amongst the $L$ masses.
While the signature of the condensation transition is manifest in the 
single-site mass distribution $p(m)$, a more direct and somewhat
natural probe for the condensation is the study of the statistics of the
maximal mass in the system.
This is because in the condensed phase, the site
where condensation occurs is also the site carrying the maximal mass. It is
from this perspective, that of the  extreme value statistics
of the masses, that we study
condensation in this work.

\section{Condensation and extreme value statistics}
In this paper we ask the question---what is the distribution of the largest
mass in the system? Thus rather than the single-site mass distribution
$p(m,M,L)$ studied previously, we consider the probability $Q(m,M,L)$ that the
largest mass in the system (of size $L$ and total mass $M$) is less than or
equal to $m$, which is an example of an extreme value distribution.

In the zero-range process the asymptotic size of the largest mass $m^*$ has
been considered by Jeon et al \cite{JMP00}. In that work the case
(\ref{fasymp}) was considered with $\gamma >3$ and it was shown that in the
fluid phase $m^* = O(\ln L)$, at the critical density $m^* = O(
L^{1/(\gamma-1)})$, and in the condensed phase $m^*/L = \rho-\rho_c$. Here we
study the statistics of the maximal mass in the more general mass transport
model described in the introduction. Moreover, we obtain the full asymptotic
probability distribution of the maximal mass for large $L$ and not just its
average size.

The theory of extreme value statistics (EVS) is well understood for the case
of independent random variables. If one considers a set of $L$ independent and
identically distributed (i.i.d) random variables, each drawn from a common
distribution $f(m)$ then the limiting distribution of their maximum,
appropriately centred and scaled, has three possible forms~\cite{Gumbel}. The
limiting distribution is: a Gumbel distribution when $m$ unbounded and $f(m)$
decays faster than a power law for large $m$; a Fr\'echet distribution when
$m$ is unbounded and $f(m)$ decays as a power law for large $m$, and a Weibull
distribution when $m$ is bounded.

In our case, however, the $L$ masses $m_i$ are correlated due to the global
constraint of conserved total mass, explicitly manifest in the delta function
in (\ref{prodm1}). Had this delta function constraint been not there, since
$f(m)\sim A m^{-\gamma}$, one would expect a Fr\'echet distribution for the
maximal mass based on the EVS of i.i.d. random variables. Here, we find that
the delta function induces nontrivial correlations that modify this naive
expectation based on the EVS of i.i.d. random variables. 

We shall show that the three phases (respectively for $\rho<\rho_c$,
$\rho=\rho_c$ and $\rho>\rho_c$) exhibit distinct forms for $Q(m,M,L)$, the
cumulative probability distribution for the largest mass. In the fluid phase
($\rho<\rho_c)$ $Q(m,M,L)$, appropriately centred and scaled, is a Gumbel
distribution; at the critical point ($\rho=\rho_c$) it becomes a Fr\'echet
distribution; in the condensed phase ($\rho>\rho_c$) the distribution of the
largest mass in the system is given by $L p_{\rm cond}(m,M,L)$ where $p_{\rm
  cond}(m,M,L)$ was computed explicitly in \cite{MEZ05,EMZ06}. In the latter
case we see that maximal mass distribution is essentially given by the tail of the single-site mass distribution. The multiplying factor $L$ simply denotes the fact
that any of the $L$ sites can carry the maximal mass. Thus, in the condensed
phase, the maximal mass distribution is neither of the three types (Gumbel,
Fr\'echet or Weibull) that one encounters in the EVS of i.i.d random
variables, indicating strong effects of the correlations between the masses.
Thus the correlation arising from the global mass conservation modifies the
maximal mass distribution both in the fluid phase ($\rho<\rho_c$) as well as
in the condensed phase ($\rho>\rho_c$). Interestingly exactly at the critical
point $\rho=\rho_c$, the maximal mass distribution is Fr\'echet as would be
prediction based on i.i.d variables indicating that the correlation is somehow
least effective exactly at the critical point.

\section{Computation of the distribution of the largest mass}

We begin by defining the cumulative distribution $Q(x,M,L)$, the probability
that the largest mass is less than or equal to $x$:
\begin{equation}
Q(x,M,L)=  \frac{I(x,M,L)}{Z(M,L)}\;.
\label{Qdef}
\end{equation}
Here,
\begin{equation}
I(x,M,L)= \prod_{i=1}^L \int_0^x {\rm d} m_i f(m_i)\,
\delta(\sum_{j=1}^L m_j -M)
\label{Idef}
\end{equation}
and $Z(M,L)$ (\ref{Zcan}) is given by
$Z(M,L) = I(\infty, M, L)$.
Equation (\ref{Qdef}) follows from the fact
that if the largest mass is $\leq x$, all masses must be $\leq x$.
We shall also consider the probability density of the largest mass
\begin{eqnarray}
P(x,M,L) &=& \frac{\partial}{\partial x} Q(x,M,L)\label{Px}\\
&=& L f(x)\frac{I(x,M-x,L-1)}{Z(M,L)}\;. \label{Px2}
\end{eqnarray}
Expression (\ref{Px2}) may  be compared to the single site mass distribution
studied in \cite{MEZ05,EMZ06}
\begin{equation}
p(m,M,L) = f(m) \frac{Z(M-m,L-1)}{Z(M,L)}\;.
\end{equation}

The Laplace transform of $I(x,M,L)$ is easy to compute from (\ref{Idef})
\begin{equation}
\int_{0}^{\infty}{\rm d} M\ I(x,M,L)e^{-sM} 
= \left[\int_0^x {\rm d} m\ f(m) e^{-sm}  \right]^L,
\end{equation}
therefore
\begin{equation}
I(x,M,L) = 
\int_{c -i \infty}^{c + \infty} 
\frac{{\rm d}s}{2\pi i}
\exp L \left\{ \rho s + \ln \left[ 
\int_0^x {\rm d} m\ f(m) e^{-sm}  \right] \right\}\;
\label{I1}
\end{equation}
where $M=\rho L$ and $c$ is chosen so that integration contour is to the right of any
singularity of the integrand.
Our task now  is to evaluate the integral (\ref{I1}). 
In \cite{MEZ05,EMZ06} the partition function
$Z(M,L) = I(\infty, M, L)$ was evaluated and we shall follow a similar
approach here.

\subsection{Fluid phase $\rho < \rho_c$}
First we consider the fluid phase in which case 
$I(x, M, L)$ may be evaluated by the saddle-point method.
Let us define
\begin{equation}
\mu_n(x,s) = \int_x^{\infty} {\rm d}m \ f(m)\ m^n\  e^{-sm}
\label{mudef}
\end{equation}
where due to our choice of normalisation in (\ref{fasymp}),
$\mu_0(0,0)=1$.

For $0< s< \infty$ and $x$ large, $\mu_n(x,s)$
is asymptotically
\begin{equation}
\mu_n(x,s) = \frac{f(x) x^n e^{-sx}}{s} \left[1 + O(1/x)\right].
\label{muna}
\end{equation}
On the other hand, for $s=0$ and $f(m)\simeq A m^{-\gamma}$ for large $m$,
we have for large $x$,
\begin{equation}
\mu_n(x,0) \simeq \frac{A}{\gamma -n -1} x^{n-\gamma+1}
\qquad \mbox{for}\quad \gamma > n+1.
\label{munb}
\end{equation}

Next, using
$\int_0^{x} dm f(m) e^{-sm}= \mu_0(0,s)-\mu_0(x,s)$, 
we may write  (\ref{I1}), for large $x$, 
as
\begin{equation}
I(x,M,L) \simeq
\int_{c -i \infty}^{c + \infty} 
\frac{{\rm d}s}{2\pi i}
\exp L \left\{ \rho s + \ln \mu_0(0,s) - \frac{\mu_0(x,s)}{\mu_0(0,s)}
+ \cdots \right\}
\label{I2}
\end{equation}
and to leading order the 
saddle point, $s_0$, of the integrand  is independent of $x$
and is given by the equation
\begin{equation}
\rho = \frac{\mu_1(0,s_0)}{\mu_0(0,s_0)}\;,
\label{sad}
\end{equation}
where we have used $\displaystyle \frac{\partial}{\partial s}\left[ \mu_0(0,s)\right]
= - \mu_1(0,s)$.
The saddle point exists and lies on the positive real axis
when $\rho<\rho_c$ given from (\ref{rhoc}) by
\begin{equation}
\rho_c = \mu_1(0,s)\;.
\end{equation}
Therefore, in the fluid phase ($\rho<\rho_c$), to leading order,
we obtain
\begin{eqnarray}
Q(x,M,L)&\simeq&
\exp\left[ -L \frac{\mu_0(x,s_0)}{\mu_0(0,s_0)}\right]
\simeq
\exp\left\{ - L \frac{A  e^{-s_0x}}{x^{\gamma} s g(s_0)}\right\}
\end{eqnarray}
where $g(s_0) = \mu_0(0,s_0)$,
and we have used (\ref{muna}).
For $x =  \frac{\ln L}{s_0} + O(\ln \ln L)$ this may be written in the
familiar form of a Gumbel distribution
\begin{equation}
Q(x,M,L)\simeq
 \exp\left\{-\exp \left[ - \left( \frac{x-a_L}{b_L} \right) \right] \right\}
\label{gumbel}
\end{equation}
where
\begin{equation}
a_L  = \frac{1}{s_0} \ln \left[ \frac{L A}{(\ln L)^{\gamma}  s_0^{1-\gamma} g(s_0) }
\right]
\quad ; \quad b_L= 1/s_0\;.
\end{equation}
Thus the probability density of the maximal mass $P(x,M,L)$ has a peak around 
$x=a_L$ with a width $b_L$. The scaling form of the distribution
around this peak has the Gumbel form (\ref{gumbel}) and implies that the
largest mass is $O(\ln L)$.

\subsection{Critical density $\rho=\rho_c$}
At the critical density $\rho=\rho_c$ the saddle point
of the integral is at $s_0=0$
and $\rho_c$ is given by (\ref{rhoc}).
In this case, 
we use (\ref{munb})
\begin{equation}
\mu_0(x,0) \simeq  \frac{A}{\gamma -1} \frac{1}{x^{\gamma-1}} 
\label{rhoc2}
\end{equation}
and find, to leading order,
\begin{eqnarray}
Q(x,M,L)&\simeq& \exp\left\{ -\frac{A}{g(0)(\gamma-1)}Lx^{-(\gamma-1)}\right\}\;.
\end{eqnarray}
This is a Fr\'echet distribution and implies that the largest mass
is $O(L^{1/(\gamma-1)})$.

\subsection{Condensed phase $\rho > \rho_c$}
We now turn to the condensed phase  where $\rho > \rho_c$. In this 
case the saddle-point method can no longer be used to evaluate
(\ref{I1}) as there is no longer a solution to (\ref{sad}).

It turns out to be convenient, in this case, to work directly with the
probability density of the largest mass $P(x,M,L)$ in (\ref{Px2}), rather than
the cumulative distribution $Q(x,M,L)$. In fact, the calculation of
(\ref{Px2}) reduces to precisely that of the distribution $p_{cond}(m,M,L)$
detailed in \cite{MEZ05,EMZ06} which we now review.

To analyse (\ref{Px2}), we need 
to consider  both $I(x,M-x,L-1)$ (in the numerator) and also
$Z(M,L)$ (in the denominator). 

First, the behaviour of $Z(M,L)=I(\infty,M,L)$ for $\rho > \rho_c$
has been computed in \cite{MEZ05,EMZ06}. The integral (\ref{I1})
\begin{eqnarray}
I(\infty,M,L)
&= & 
\int_{-i \infty}^{i\infty} 
\frac{{\rm d}s}{2\pi i} \exp L
\left[ \rho s + \ln \mu_0(0,s)\right]
\end{eqnarray}
will be  dominated by small values of  $s$. Therefore, one expands
the  term $\ln \mu_0(0,s)$  for $s$ small. The expansion
depends on the value  of $\gamma$ and is given by
\begin{eqnarray}
\mbox{For}\quad 3< \gamma\;,\quad
\ln \mu_0(0,s) &\simeq&  - s \rho_c + b s^{\gamma-1} + \cdots 
\label{lm1}\\
\mbox{For}\quad 2< \gamma <3\;,  \quad
\ln \mu_0(0,s) &\simeq&  - s\rho_c  + \frac{\Delta}{2} s^{2}
+ \cdots  + b s^{\gamma-1}+\cdots 
\label{lm2}
\end{eqnarray}
where $b = A \Gamma(1-\gamma)$, $\Delta^2 =
(\mu_2(0,0)-\mu_1^2(0,0))/\mu_0(0,0)$ and $\rho_c$ is given by (\ref{rhoc2}).

Similarly, we consider the numerator in (\ref{Px2}), for which the relevant
integral (\ref{I1}) is
\begin{eqnarray}
I(x,M-x,L-1)
&= & 
\int_{-i \infty}^{i\infty} 
\frac{{\rm d}s}{2\pi i} \frac{\exp L H(s,x)}{\mu_0(0,s)}
\label{I3}
\end{eqnarray}
where  for large $x$ (we will be considering $x= O(L)$ in the large $L$ limit)
\begin{equation}
H(s,x) \simeq 
\rho s  -xs/L  + \ln \mu_0(0,s) \;.
\label{H(x)}
\end{equation}
We now expand this expression for
$H(s,x)$ for small $s$, using (\ref{lm1},\ref{lm2}).

As a result of the small large $L$, small $s$ expansions just described, one obtains from (\ref{Px2})
\begin{equation}
P(x,M,L) \simeq {L f(x)}
\frac{ W_L (x/L  -(\rho -\rho_c))}
{W_L ( -(\rho-\rho_c))} \label{PW}
 \end{equation}
where $W_L(y)$ has different forms according
to the cases $\gamma >3$ and $2<\gamma<3$
which we  discuss separately below.

We note that (\ref{PW}) yields the same leading large $x$ behaviour for
$P(x,M,L)/L$ as for the single site mass distribution $p(x,M,L)$ computed in
\cite{EMZ06}. This is due to the fact that for $x \gg \ln L$ the same leading
behaviour is obtained for $I(x,M-x,L-1)$ as for $I(\infty,M-x,L-1)$. Moreover,
we can identify the peak in the probability density of the largest mass with
the peak $p_{cond}$ in the single site distribution.\\

\noindent {\bf Case (i)  $\gamma > 3$}\\
In this case $W_L(y)$ in (\ref{PW}) is given by
\begin{equation}
W_L(y) = \int_{-i \infty}^{i \infty} \frac{{\rm d}s}{2\pi i}
\exp L \left[ -y s + \frac{\Delta}{2} s^{2}/2
+ \cdots +  b s^{\gamma-1}+\cdots \right]\;.
\end{equation}
The important behaviour of the function $W_L(y)$ determined in \cite{MEZ05,EMZ06} 
may be summarised as
\begin{eqnarray}
W_L(y) &\simeq& \frac{1}{\sqrt{2\pi \Delta^2 L}} e^{-y^2L/2 \Delta^2}
\qquad \mbox{for} \quad y = o(L^{-1/3})\\
W_L(y) &\simeq& \frac{A}{ |y|^\gamma L^{\gamma-1}}
\qquad \mbox{for} \quad y = O(1)\;,
\end{eqnarray}
from which we deduce that for $\rho > \rho_c$
\begin{equation}
P(x,M,L) \simeq 
\left(\frac{L(\rho_c-\rho)}{x}\right)^{\gamma}
 W_L (x/L -(\rho -\rho_c))\;.
\end{equation}
Thus the probability density of the maximal mass is peaked around
$x=(\rho-\rho_c)L$ and near this peak, that is,
over a scale of $x- L(\rho -\rho_c) = 
o(L^{2/3}$), the probability density
has a normalized gaussian form
\begin{equation}
P(x,M=\rho L, L)\simeq \frac{1}{\sqrt{2\pi \Delta^2 L}}\, 
\exp\left[-\frac{{\left(x-(\rho-\rho_c)L\right)}^2}{2\Delta^2 L}\right].
\end{equation}
\vspace*{2ex}
\noindent {\bf  Case (ii) $3> \gamma > 2$}\\
From (\ref{Px2}) 
one obtains (\ref{PW})
where now  
\begin{equation}
W_L(y) = L^{-1/(\gamma-1)} 
V_{\gamma}\left[ L^{(\gamma-2)/(\gamma-1)}y\right]
\label{Vg}
\end{equation}
and
\begin{equation}
V_\gamma(z) = \int_{-i \infty}^{i \infty} \frac{{\rm d}s}{2\pi i}
e^{-zs + bs^{\gamma-1}}\;
\end{equation}
with, as before, $b= A \Gamma(1-\gamma)$.
The function $V_{\gamma}(z)$ is highly asymmetric around $z=0$
and has the following asymptotic behaviour~\cite{MEZ05,EMZ06}
\begin{eqnarray}
V_{\gamma}(z) &\simeq & \frac{A}{|z|^{\gamma}}\quad {\rm as}\quad z\to -\infty
\label{Vga}
 \\
&\simeq & c_1 z^{(3-\gamma)/{2(\gamma-2)}}\, \exp[-c_2 z^{(\gamma-1)/(\gamma-2)}]\quad {\rm 
as}\quad z\to \infty
\end{eqnarray}
where $c_1$ and $c_2$ are two constants that were computed in \cite{EMZ06}
\begin{eqnarray}
c_1 &=& \left[ 2 \pi(\gamma-2)(b(\gamma-1))^{1/(\gamma-2)}\right]^{-1/2}\;,\\
c_2 &=& (\gamma-2) \left[ (\gamma-1)(b(\gamma-1))^{1/(\gamma-2)}\right]^{-1}\;.
\end{eqnarray}  
Using (\ref{Vg}) and  (\ref{Vga}) in (\ref{PW}),
we obtain the normalized maximal mass density 
\begin{equation}
P(x,M,L) \simeq \frac{1}{L^{1/(\gamma-1)}}
V_{\gamma}\left[ \frac{x- L(\rho-\rho_c)}{L^{1/(\gamma-1)}}\right]\;.
\end{equation}
Thus, in the condensed phase ($\rho>\rho_c$) the maximal mass distribution
has neither of the three forms associated with the EVS of i.i.d random variables.
Rather, the scaled distribution has a gaussian form for $\gamma>3$ and has
a nontrivial non-gaussian form for $2<\gamma<3$. In both cases the largest
mass is $O(L)$.

\section{Discussion}
In summary, we have probed the condensation transition in a generalized class
of mass transport models by studying the cumulative probability distribution
of the maximal mass in the system. The masses in this model are globally
constrained via the mass conservation law. Without this global constraint the
masses would have been independent random variables each drawn from a
power-law distribution and based on the EVS of the i.i.d random variables one
would expect a Fr\'echet distribution for the maximal mass for all densities
$\rho$. Instead, we have shown via exact asymptotic calculation that the
global constraint is sufficiently strong to modify this expectation based on
i.i.d. variables and one obtains respectively a Gumbel distribution (for
$\rho<\rho_c$), a Fr\'echet distribution (for $\rho=\rho_c$) and a completely
different distribution in the condensed phase (for $\rho>\rho_c$). In the
latter case, the distribution is gaussian for $\gamma>3$ and highly
non-gaussian for $2<\gamma<3$. Thus, our exact 
results for this model form a useful
addition to the list of exactly solvable cases for the distribution of extreme
values of a set of {\em correlated} random variables, for example, the
eigenvalues of random matrices ~\cite{Edelman,TW,Johansson,Johnstone,BBP,MBL},
energies of configurations a directed polymer in a random medium~\cite{DM01},
the heights in  a one-dimensional interface~\cite{G1,MC,SM,GMOR}.
We note that the Gumbel distribution emerging out of a global constraint
was also observed recently in the context of complex chaotic states \cite{LTBM}.

We have also seen how in the condensed phase the bump, $p_{\rm cond}(m)$ in
the single-site mass distribution is directly related to the probability
density of the size of largest mass in the system.

Finally, we note that as the density $\rho$ is increased the phenomenon of
condensation is manifested through changes in the asymptotic behaviour of the
large mass distribution as described above. It would be of interest to analyse
more closely the crossovers, both from the Gumbel distribution
for $\rho <\rho_c$ to the Fr\'echet
distribution at $\rho=\rho_c$
and from the  Fr\'echet
distribution at $\rho=\rho_c$ to the condensed phase distribution 
for $\rho > \rho_c$, as the density is
increased. This would involve the consideration of finite-size effects and
finite-size scaling near the transition.

\ack 
MRE thanks the CNRS for a Visiting Professorship
and the hospitality of LPTMS.
A visit of SNM to Edinburgh was
supported by  EPSRC Programme grant EP/E030173/1.\\

\noindent {\bf References}\\


\begin{thebibliography}{10}

\bibitem{EMZ04}  M. R. Evans, S.N. Majumdar, and R. K. P. Zia, J. Phys. A: Math.
Gen {\bf 37} (2004) L275

\bibitem{OEC}  O.J.~O'Loan, M.R.~Evans and M.E.~Cates, Phys. Rev. E {\bf 58}%
, 1404 (1998).

\bibitem{KMH05}  J Kaupuzs, R Mahnke, RJ Harris,
Phys. Rev. E {\bf 72}, 056125 (2005)

\bibitem{granular}  
D. van der Meer, K. van der Weele, P. Reimann and D. Lohse
J. Stat. Mech.: Theor. Exp., P07021  (2007); J. Torok. 
Physica A {\bf 355} 374 (2005).

\bibitem{KLMST}  Y.~Kafri, E.~Levine, D.~Mukamel, G.M.~Sch\"{u}tz and J.~T{%
\"{o}}r{\"{o}}k, Phys.~Rev.~Lett. {\bf 89}, 035702 (2002).


\bibitem{DM}
S. N. Dorogovtsev and J.F.F. Mendes,
 {\em Evolution of Networks} (OUP, Oxford, 2003)

\bibitem{AELM}
A. G. Angel, M. R. Evans, E. Levine, D. Mukamel
Phys. Rev. E, {\bf 72}, 046132 (2005)

\bibitem{AHE06}
A. G. Angel, T. Hanney, and M. R. Evans,
Phys. Rev. E {\bf 73}, 016105 (2006)


\bibitem{CLMNW}  S.N. Coppersmith, C.-h. Liu, S. Majumdar, O. Narayan, T.A.
Witten, Phys. Rev. E., {\bf 53}, 4673 (1996).

\bibitem{MKB}  S.N.~Majumdar, S.~Krishnamurthy and M.~Barma, Phys. Rev.
Lett. {\bf 81}, 3691 (1998); J. Stat. Phys. {\bf 99}, 1 (2000).

\bibitem{ARAP}  J. Krug and J. Garcia, J. Stat. Phys., {\bf 99} 31 (2000);
 R. Rajesh and S. N. Majumdar, J. Stat. Phys., {\bf 99} 943 (2000).


\bibitem{Bertin} E. Bertin, J. Phys. A: Math. Gen. {\bf 39}, 1539 (2006).

\bibitem{EH05}  M. R. Evans and T. Hanney, J. Phys. A: Math. Gen. {\bf 38}, R195 (2005)



\bibitem{ZS} F. Zielen and A. Schadschneider, Phys. Rev. Lett {\bf 89}
090601 (2002)



\bibitem{ZEM04}
 R. K. P. Zia, M. R. Evans, S.N. Majumdar, J.
Stat. Mech. : Theor. Exp. (2004) L10001.



\bibitem{EMZ06a}  M. R. Evans, S.N. Majumdar, and R. K. P. Zia,
J. Phys. A: Math. Gen {\bf 39}, 4859-4873 (2006)

\bibitem{queue}
J. Walrand {\em An Introduction to Queueing Networks} (Prentice-Hall, 1988)


\bibitem{BBJ}  P. Bialas, Z. Burda, and D. Johnston, Nucl. Phys. B {\bf 493}%
, 505 (1997).


\bibitem{Evans00}  M.R.~Evans, Braz. J. Phys. {\bf 30}, 42 (2000).

\bibitem{JMP00}
I. Jeon, P. March and B. Pittel, Ann. Prob. {\bf 28} 1162 (2000)


\bibitem{GSS03}  S.~Gro\ss kinsky, G.M Sch\"{u}tz and H.~Spohn, J. Stat. Phys 
{\bf 113}, 389 (2003).

\bibitem{Godreche03}  C.~Godr\`{e}che, J. Phys. A: Math. Gen., {\bf 36}, 6313
(2003).

\bibitem{MEZ05}  S.N. Majumdar, M. R. Evans, and  R. K. P. Zia,
Phys. Rev. Lett. {\bf 94}, 180601 (2005)


\bibitem{EMZ06} M. R. Evans, S.N. Majumdar, and R. K. P. Zia, J. Stat. Phys. 
{\bf 123} 357 (2006)

\bibitem{FLV07}
P. A. Ferrari, C. Landim,  and V. V. Sisko, J. Stat. Phys
{\bf 128}  1153 (2007)



\bibitem{Gumbel} E.J. Gumbel, {\em Statistics of Extremes} (Columbia University, New 
York, 1958).





\bibitem{Edelman} A. Edelman, Siam J. Matrix Anal. Appl. {\bf 9}, 543 (1988).

\bibitem{TW} C.A. Tracy and H. Widom, Commun. Math. Phys. {\bf 159}, 151 (1994).

\bibitem{Johansson} K. Johansson, Comm. Math. Phys. {\bf 209}, 437 (2000).

\bibitem{Johnstone} I.M. Johnstone, Ann. Stat. {\bf 29}, 295 (2001).

\bibitem{BBP} G. Biroli, J.-P. Bouchaud, and M. Potters, Europhys. Lett. {\bf 78}, 10001 
(2007).

\bibitem{MBL} S.N. Majumdar, O. Bohigas, and A. Lakshminarayan, J. Stat. Phys. {\bf 131}, 
33 (2008).


\bibitem{DM01} S.N. Majumdar and P.L. Krapivsky, Phys. Rev. E {\bf 62}, 7735
  (2000); D.S. Dean and S.N. Majumdar, Phys. Rev. E {\bf 64}, 046121 (2001).

\bibitem{G1} G. Gyorgyi et. al. Phys. Rev. E {\bf 68}, 056116 (2003).

\bibitem{MC} S.N. Majumdar and A. Comtet, Phys. Rev. Lett. {\bf 92}, 225501 (2004);
J. Stat. Phys. {\bf 119}, 777 (2005).

\bibitem{SM} G. Schehr and S.N. Majumdar, Phys. Rev. E {\bf 73}, 056103 (2006).


\bibitem{GMOR} G. Gyorgyi, N.R. Maloney, K. Ozogany, and Z. Racz, Phys. Rev. E {\bf 75}, 
021123 (2007).


\bibitem{LTBM} A. Lakshminarayan, S. Tomsovic, O. Bohigas, and S. N. Majumdar, 
 Phys. Rev. Lett. {\bf 100}, 044103 (2008).




\end{thebibliography}
\end{document}